\documentclass[reprint,amsmath,amssymb,aps,groupedaddress]{revtex4-1}
\usepackage{graphicx}
\usepackage{dcolumn}
\usepackage{bm}
\begin{document}
\newcommand{\revtex}{REV\TeX\ }
\newcommand{\classoption}[1]{\texttt{#1}}
\newcommand{\macro}[1]{\texttt{\textbackslash#1}}
\newcommand{\m}[1]{\macro{#1}}
\newcommand{\env}[1]{\texttt{#1}}
\setlength{\textheight}{9.5in}
\bibliographystyle{apsrev4-1}
\preprint{APS/123-QED}

\title{Coexisting two-dimensional electron and hole gases highly confined at the interfaces of undoped KTaO$_3$-sandwiching heterostructures}
\author{Ning Wu}
\affiliation{Beijing National Laboratory for Condensed Matter Physics, Institute of Physics, Chinese Academy of Sciences, Beijing 100190, China}
\affiliation{School of Physical Sciences, University of Chinese Academy of Sciences, Beijing 100190, China}
\author{Bang-Gui Liu}
\email[]{Email: bgliu@iphy.ac.cn}
\affiliation{Beijing National Laboratory for Condensed Matter Physics, Institute of Physics, Chinese Academy of Sciences, Beijing 100190, China}
\affiliation{School of Physical Sciences, University of Chinese Academy of Sciences, Beijing 100190, China}
\date{\today}

\begin{abstract}
Two-dimensional electron gas (2DEG) in interfaces and surfaces based on perovskite SrTiO$_3$ (STO) has exhibited various interesting phenomena and is used to develop oxide electronics. Recently, KTaO$_3$ (KTO) shows great potential and is believed to host more exciting effects and phenomena toward novel devices. Here, through first-principles investigation and analysis, we find two types of coexisting 2DEG and 2D hole gas (2DHG) highly confined at the interfaces in undoped STO/KTO/BaTiO$_3$ heterostructures, when the KTO thickness $m$ reaches a crititcal value. The two interfaces are made by (SrO)$^0$/(TaO$_2$)$^+$  and (KO)$^-$/(TiO$_2$)$^0$  for the A-type, and by (TiO$_2$)$^0$/(KO)$^-$ and (TaO$_2$)$^+$/(BO)$^0$ for the B-type. The 2D electron carriers originate from Ta-$5 d_{xy}$ states at the interface including TaO$_2$ atomic layer, and the hole carriers from O-$2 p_x/p_y$ orbitals at the other interface. The electron and hole effective masses are 0.3$m_0$ and $1.06\sim 1.12 m_0$, respectively, where $m_0$ is mass of free electron, and the 2D carrier concentrations are in the order of $10 ^{13}$ cm$^{-2}$. Our analysis indicates that the interfacial 2DEG and 2DHG are simultaneously formed because of the band bending due to the polar discontinuity at the interfaces and the stress-induced polarization within the KTO layer. These could stimulate more exploration for new phenomena and novel devices.
\end{abstract}

\maketitle

\section{Introduction}

Since the two-dimensional electron gas (2DEG) at the LaAlO$_{3}$ (LAO)/SrTiO$_{3}$ (STO) interfaces was discovered by Ohtomo and Hwang\cite{ohtomo2004high}, great attention has been paid to the study of various oxide interfaces for their useful properties\cite{nakagawa2006some}. Although the 2DEG has been widely observed, it is very much challenging to achieve two-dimensional hole gas (2DHG) because p-type interfaces tend to be insulating\cite{nakagawa2006some}. Until recently, the coexistence of 2DEG and 2DHG in epitaxially grown STO/LAO/STO was shown with electrical transport measurements\cite{lee2018direct}. Coexisting 2DEG and 2DHG in the oxide interfaces enables an exploration of the exciting new physics of confined electron-hole systems\cite{lee2018direct}, including long-lifetime bilayer excitons\cite{millis2010electron}, Coulomb drag with spin-orbit coupling\cite{tse2007coulomb}, bilayer electron-hole superconductivity and the Bose-Einstein condensation of excitons\cite{eisenstein2004bose}. Albeit the 2DEG was usually considered to be confined at the interface, the spatial extent of the electron gas has been found to vary from a depth of a few nanometers to hundreds of micrometers\cite{basletic2008mapping}, with the 2DEG in the LAO/STO interface spreading 2.5 nm into in the STO substrate\cite{shanavas2016theoretical}. These issues can be attributed to many factors, including oxygen vacancy concentration\cite{basletic2008mapping}, temperature of the system\cite{siemons2007origin}, charge density of the electron gas\cite{khalsa2012theory}, cationic exchange\cite{takizawa2006}, and ionic relaxation at the interface\cite{okamoto2006,2009avoiding}.

For perovskite KTaO$_3$ (KTO), some excellent properties have been observed, such as high mobilities and signatures of spin-orbit coupling in low-temperature magnetoresistance data\cite{zhang2019unusual}, spin-charge conversion of 2DEG at KTO surface\cite{vicente2021spin,zhang2019thermal}, spin-polarized 2DEGs at a magnetic EuO/KTO interface\cite{2018high}, and 2D superconductivity discovered at the KTO-based oxide interfaces\cite{2021critical,2021two,chen2021two,ma2020superconductor,chen2021electric}  (see \cite{add4} for an overall review). In view of realistic applications, a main challenge is the epitaxial growth of KTO because of very reactive potassium atoms. One must take into account the polar character of KTO layers\cite{2012subband,2019structure}. Annealing in vacuum allows the formation of isolated oxygen vacancies, followed by a complete rearrangement of the top layers resulting in an ordered pattern of KO and TaO$_2$ stripes, leaving approximately half (KO)$^-$ coverage on the (TaO$_2$)$^+$ surface\cite{2018polarity}. Defect-free surfaces were only achieved for samples cleaved below room temperature. As a consequence, it is difficult to obtain a single-terminated  surface for the KTO (001) due to its polar nature. However, the polar nature of KTO (001) can recover by a deposited capping layer, and actually the polarization of KTO (001) has been realized by simple deposition of Al metal onto KTO single crystals\cite{vicente2021spin}.

Here, we construct a KTO-sandwiching heterostructure by putting an epitaxial KTO layer on STO(001) substrate and capping the KTO surface by a BaTiO$_3$(BTO) layer, making a stoichiometric STO/KTO/BTO heterostructure, and then optimize its structure and investigate its electronic properties through first-principles calculation and analysis. Our study show that there are two pairs of interfaces in the heterostructure due to two different terminations, and accordingly two pairs of 2DEG and 2DHG are formed and highly confined at the interfaces. Furthermore, it is shown that the 2DEG consists of Ta-$5 d_{xy}$ states and the 2DHG O-$2 p_x/p_y$ states, which can be attributed to intrinsic electric field in the polar KTO layer. More detailed results will be presented in the following.

\section{COMPUTATIONAL DETAILS}

The structural optimization and electronic properties are studied within the density-functional thoery, as implemented in the Vienna Ab initio Simulation Package (VASP)\cite{kresse1999ultrasoft,blochl1994projector}. To describe the exchange-correlation functional, we choose the general gradient approximation (GGA) with the Perdew-Burke-Ernkzerhof for solids (PBEsol) parametrization\cite{perdew1996generalized,perdew2008restoring}. The on-site Coulomb interaction in the 5d states of transition-metal ions is corrected by using the DFT$+$U method\cite{anisimov1997first}, where U is the Hubbard parameter. The $U_{\rm eff}$ = 3 eV is used for Ta 5d state\cite{Zhang2018Strain}, and 4.36 eV for Ti 3d states\cite{add3}, because it is well established that the $U_{\rm eff}$ values are good to describing these strongly-correlated electronic states, in addition to excellent bulk lattice constants, for STO/BTO and KTO\cite{add3,Zhang2018Strain}. A Monkhorst-Pack k-point grid of ($4\times4\times1$) is used for the reciprocal space integration and the plane wave energy cutoff is set to 500 eV. Our convergence standard requires that the Hellmann-Feynman force on each atom is less than 0.01 eV/{\AA} and the absolute total energy difference between two successive loops is smaller than $1\times10^{-5}$ eV. For the k-point grid of KTO bulk calculations, we use $11\times11\times11$ for structural optimization, and $15\times15\times15$ for electric polarization.

\section{RESULTS AND DISCUSSION}

\subsection{Structures and optimization}

We investigate stoichiometric STO/KTO/BTO sandwich heterostructure to model a BTO-capped epitaxial KTO (001) ultrathin film on STO(001) substrate. The non-periodic STO/KTO/BTO heterostructure is simulated with a periodic sandwich slab model with a vacuum layer (20 \AA{}). To reflect epitaxial KTO growth on STO (001) substrate (and BTO on KTO), the in-plane lattice constant $a$ of the heterostructure is constrained to that of the experimental equilibrium bulk (cubic) STO, $a_{\rm STO} = 3.905$. The atomic coordinates and out-of-plane lattice constant are fully optimized. We note that the charge states in the KTO are positive for the Ta$^{5+}$O$_2^{2-}$ atomic layer (AL) and negative for the K$^{1+}$O$^{2-}$ AL, but in the ionic limit the charge states in the STO (BTO) layer are neutral for Sr$^{2+}$O$^{2-}$ (Ba$^{2+}$O$^{2-}$) AL and Ti$^{4+}$O$_{2}$$^{2-}$ AL. There are two interfaces in the STO/KTO/BTO sandwich heterostructure: STO/KTO and KTO/BTO. Keeping it stoichiometric, we need to make each of the three layers have a thickness in unit cell (uc), respectively. The STO/KTO interface can take one of the two cases, (SrO)$^{0}$/(TaO$_2$)$^{+}$ and (TiO$_2$)$^{0}$/(KO)$^{-}$. Consequently, the KTO/BTO interface needs to be the corresponding one of (KO)$^{-}$/(TiO$_2$)$^{0}$ and (TaO$_2$)$^{+}$/(BO)$^{0}$. Therefore, we have two types of the STO/KTO/BTO sandwich heterostructure according to the two pairs of interfaces: (SrO)$^{0}$/(TaO$_2$)$^{+}$-(KO)$^{-}$/(TiO$_2$)$^{0}$ (A-type) and (TiO$_2$)$^{0}$/(KO)$^{-}$-(TaO$_2$)$^{+}$/(BO)$^{0}$ (B-type).

\begin{figure}[h]
\centering
\includegraphics[width=0.45\textwidth]{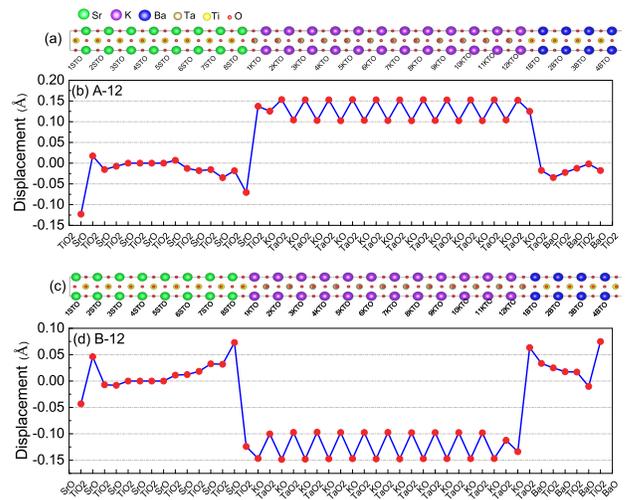}
\caption{\label{Fig1}
The side view of crystal structure (a) and AL-resolved cation displacements (b) of the A-12 optimized heterostructure, and those corresponding (c, d) of the B-12. The cation displacement is defined with respect to the nearest O anions. The Sr, Ti, K, Ta, Ba and O atoms in (a) and (c) are shown by the green, yellow, purple, brown, blue and red balls, respectively.
}
\end{figure}

For our computational investigation, we set the thickness of STO layer to 8 uc to effectively simulate the STO substrate, and use 4 uc for the BTO capping layer. In contrast, we vary the thickness of KTO layer ($m$  in uc) to consider different thickness of KTO layer. The structure of the central 3 uc in the STO layer is fixed at the STO bulk  to reflect the bulk behavior in the interior of the STO substrate. The 20 {\AA} of vacuum is enough to avoid artificial effects in the slab model. We use A-$m$ to denote the A-type STO/KTO($m$)/BTO heterostructure with the thickness $m$ uc for the KTO layer, and take B-$m$ for the B-type one. At the STO/KTO interface, the STO layer is terminated with SrO AL and the KTO layer with TaO$_2$ AL for the A-$m$ heterostructure,  the STO layer is terminated with TiO$_2$ AL and the KTO layer with KO AL for the B-$m$ heterostructure. Accordingly, the KTO/BTO interface is KO/TiO$_2$ (top BTO surface is a BaO AL), and TaO$_2$/BaO (top BTO surface is a TiO$_2$ AL), respectively. These will make big differences in electronic properties. We have completed structural optimization with  $m$ = 4, 6, 8, 10,  and  12 uc.

The optimized atomic structures for A-12 and B-12 heterostructures are shown in Figs. 1(a) and 1(c), respectively. To show their structural features, we present the AL-resolved out-of-plane cation displacements with respect to the centers of the nearest O anions in Figs. 1(b) and 1(d) for A- and B-12 heterostructures, respectively. In the KTO layer, the value of displacement $d$ is positive for A-12 heterostructure, but negative for B-12 heterostructure. We also calculate spontaneous polarization of optimized tetragonal bulk KTO with the in-plane lattice constants $a$ set to 3.905\AA{}, obtaining 50 $\mu$C/cm$^2$ (0.476$e/a^2$, where $e$ is electron charge), which means a ferroelectric phase for stressed KTO\cite{add5}. These imply that within the KTO layer, the displacement $d$ can reflect out-of-plane polarization and the opposite $d$ values between the A-12 and B-12 are caused by the two different types of atomic terminations. Moreover, when resolved to ALs in the KTO, the absolute displacement $d_1$ for the KO AL is less than that $d_2$ for the TaO$_2$ AL, $d_1 < d_2$. It is noticed that the displacements $d$ in KTO thin film is substantially larger than those in STO and BTO layers, which implies that spontaneous polarization is very small in the BTO and STO layers. This is consistent with the intrinsic nonpolar state in STO and the stabilized nonpolarized state of BTO thin film under 5 nm\cite{fredrickson2015switchable}. On the other hand, there are some noticeable displacements at the STO or BTO side of the interfaces, which can be attributed to interfacial effects, and the displacements of the KTO layer becomes smaller at the interfaces,  indicating that the polar discontinuity caused by polar (KTO)/nonpolar (STO, BTO) interface is reduced by the structural distortion. For the other $m$ values, we obtain similar structural features.

\subsection{Interfacial 2D electron and hole gases}

\begin{figure}
\centering
\includegraphics[width=0.45\textwidth]{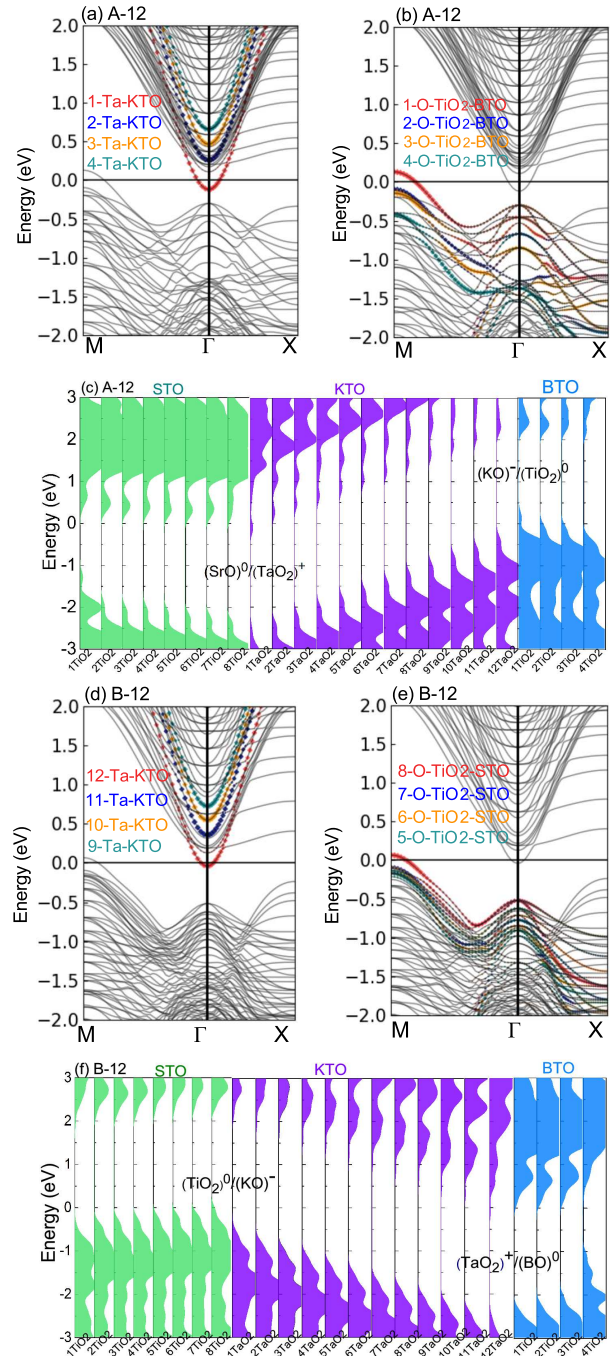}
\caption{\label{Fig2}
Orbital-resolved band structures with the Ta-5d$_{xy}$ states of TaO$_2$ ALs (a) and the O-2p$_x$/p$_y$ states of TiO$_2$ ALs (b) marked and AL-resolved DOS (c) of the A-12 STO/KTO/BTO heterostructure, and those corresponding (d, e, f) of the B-12. The AL-resolved DOS is mainly contributed by the TiO$_2$ or TaO$_2$ AL.
}\end{figure}

To display the electronic properties of the A-12 and B-12 heterostructures, we plot the orbital-resolved band structures and AL-resolved density of states (DOS) in Fig. 2. The effect of the KTO thickness $m = 4, 6, 10, 12$ on electronic bands of the STO/KTO/BTO heterostructure is presented in Fig. S1 and S2 (supporting information). It is clear that there exists an insulator-to-metal transition at $m = 6$ for both the A-type and B-type cases. It is insulating for $m = 4$ and metallic for $m = 10$ and 12. The electronic band structures are calculated along M ($\pi$, $\pi$)$\rightarrow$ $\Gamma$ ($0$, $0$) $\rightarrow$ X (0, $\pi$) direction, and the DOS plots the sum of orbital contributions from B (B = Ti, Ta) atom and O atom of the BO$_2$ ALs across the systems. We must notice an attractive feature that for the A-$m$ and B-$m$ heterostructures in metallic states, across the Fermi level are only the valence band maximum (VBM) at the $M$ point and the conduction band minimum (CBM) at $\Gamma$ point, similar to some electronic states in ultrathin KTO films. More interestingly, they are located at the different interfaces in the real space. These remain true for both A-type and B-type heterostructures. This is distinctively different from the STO system, where a multiband electronic structure was obtained for the 2DEG at surfaces or interfaces\cite{2014mixed}.

\begin{figure}
\centering
\includegraphics[width=0.42\textwidth]{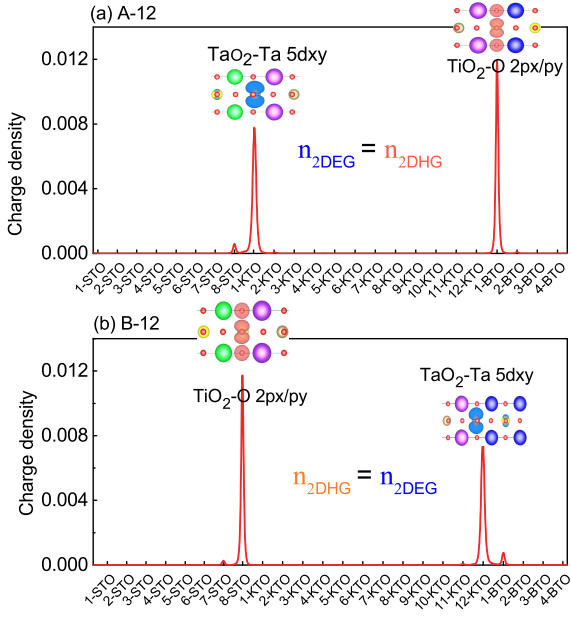}
\caption{\label{Fig3}
The monolayer-resolved charge densities of the 2DEG electrons and 2DHG holes in the A-12 (a) and B-12 (b) heterostructures. The electron part is calculated between $E_{f}$ and CBM, and the hole part between VBM and $E_{f}$. The insets show corresponding 3D charge densities with the increment 0.0005 $e$/\AA{}$^3$ ($e$ is electron charge) at the interfaces, with blue denoting the negative charge from Ta-5d$_{xy}$ and orange the positive charge from O-2p$_x$/p$_y$. }
\end{figure}

Combining the orbital-resolved band structures with the AL-resolved DOS in Fig. 2, the occupied state in the CBM is from the Ta-$5 d_{xy}$ orbital of the (SrO)$^{0}$/(TaO$_2$)$^{+}$ (STO/KTO) interface for A-type or the (TaO$_2$)$^{+}$/(BO)$^{0}$ (KTO/BTO) interface for B-type; and the unoccupied state in the VBM is mostly from the O-$2 p_{x}/p_{y}$ orbitals of the TiO$_2$ AL at the  (KO)$^{-}$/(TiO$_2$)$^{0}$ (KTO/BTO) interface for A-type or the (TiO$_2$)$^{0}$/(KO)$^{-}$ (STO/KTO) interface for B-type. Concretely, for the A-12 heterostructure, the electrons transferred from the VBM states at the STO/KTO interface are captured by the CBM (creating electron carriers) at the KTO/BTO interface, which results in unoccupied O-$2 p_{x}/p_{y}$ VBM states (creating holes) at the KTO/BTO interface. For the B-12 heterostructure, the electron carriers are contributed by the CBM states at the KTO/BTO interface and the hole carriers result from the empty VBM states at the STO/KTO interface. This implies that a pair of 2D electron and hole gases is formed at the two interfaces in the STO/KTO/BTO heterostructures. These are presented in Fig. 3, which indicates that the electron and hole carriers are highly confined to in-plane conduction at the interfaces, making a 2D electron gas (2DEG) and a 2D hole gas (2DHG). The electron carriers are separated from the hole carriers by the KTO layer, and the 2D electron concentration is equal to the 2D hole concentration, keeping the electric neutrality over the heterostructures.

For KTO(001) thin films, there are large Rashba energy splitting due to the spin-orbit coupling (SOC) in Ta-5d$_{xz}$ and 5d$_{yz}$ bands, but the SOC has little effect on Ta-5d$_{xy}$ bands\cite{zhang2019unusual,add2}. Here, the 2DEG consists of Ta-5d$_{xy}$ states and the 2DHG O-2p$_x$/p$_y$ states, and therefore the SOC effect is little on electronic structures of the 2DEG and 2DHG. Effective carrier mass ($m^{*}$) is an important parameter. It can be evaluated from the band dispersion, $m^{*} = \frac{\hbar^{2}}{d^{2}E(k)/dk^{2}}$\cite{xia2018universality} ($\hbar$ is Planck's constant and $E(k)$ describes the band). For our cases, the effective electron mass $m_{e}^{*}$ at $\Gamma$ point is $0.3 \ m_0$ ($m_0$ is the mass of the free electron). This is in excellent agreement with experimental value $0.3 \ m_0$ for the KTO surface 2DEG measured by ARPES\cite{2012subband} and nearly half that of a surface 2DEG on STO\cite{2011creation}. Moreover, the effective hole masse $m_{h}^{*}$ is anisotropic. For A-12, $m_{h}^{*}$ is $1.06 \ m_0$ along the $\Gamma-M$ direction and $1.08 \ m_0$ along the $X-M$ direction. For B-12, $m_{h}^{*}$ is $1.10 \ m_0$ along the $\Gamma-M$ direction and $1.12 \ m_0$ along the $X-M$ direction. The hole effective masses at $M$ point for A-12 and B-12 are similar with the hole effective mass of $1.2 \ m_0$ found for STO/LAO/STO structure\cite{lee2018direct}. Our calculation shows that the carrier concentration values of the 2DEG and 2DHG are $3.87 \times 10 ^{13}$ cm$^{-2}$ (0.059$/a^2$) for A-12, and  $2.23 \times 10 ^{13}$ cm$^{-2}$ (0.034$/a^2$) for B-12.

\subsection{Electrostatic potentials and electric field trend}

The monolayer-resolved planar averaged electrostatic potentials of the optimized A-12 and B-12 heterostructures are shown (black solid lines) in Figs. 4(a) and 4(c), respectively. To better elucidate the oscillating curves, they are further averaged along the z direction\cite{add1}, resulting in the smooth curves (red solid lines) in Figs. 4(a) and 4(c). For comparison, we also plot the corresponding curves (blue dash lines) before optimization (without relaxation). The most important feature is the opposite slope inside the KTO layer between the A-12 and B-12 heterostructures. For the A-12 in Fig. 4(a), there exists an ascendent slope in the electrostatic potential across the KTO layer from the (SrO)$^{0}$/(TaO$_2$)$^{+}$ (STO/KTO) interface to the (KO)$^{-}$/(TiO$_2$)$^{0}$ (KTO/BTO) interface, whereas for the B-12 in Fig. 4(c), a descendent slope in the electrostatic potential across the KTO layer from the (TiO$_2$)$^{0}$/(KO)$^{-}$ (STO/KTO) interface to the (TaO$_2$)$^{+}$/(BaO)$^{0}$ (KTO/BTO) interface. This distinctive feature can be explained by observing that the electrostatic potential always increases from the (TaO$_2$)$^+$ end to the (KO)$^-$ end of the polar KTO layer sandwiched in both A-type and B-type heterostructures.

\begin{figure}
\centering \includegraphics[width=0.4\textwidth]{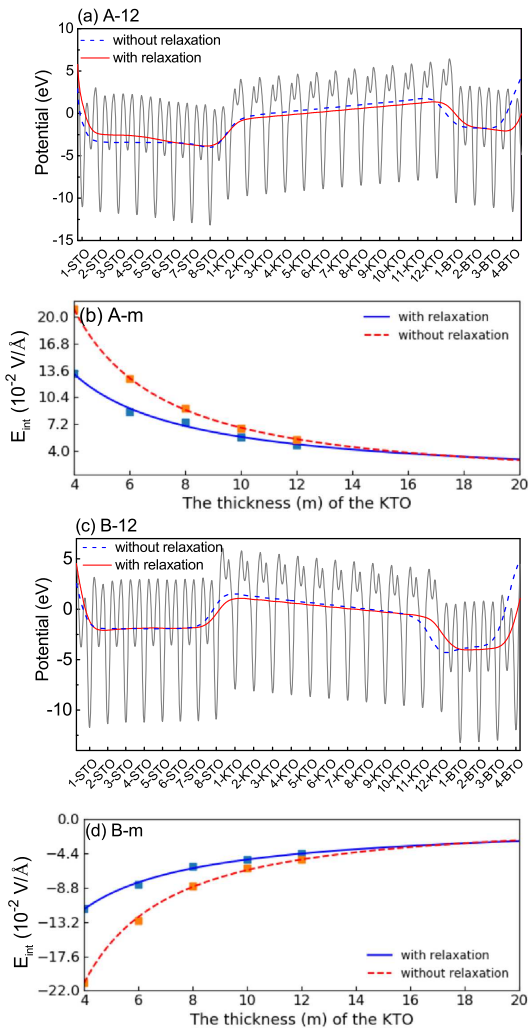}
\caption{\label{Fig4} The planar averaged electrostatic potential (oscillating gray line) and the further-averaged smooth curves without (dashed blue line) and with (solid red line) structural relaxation for the A-12 (a) and the B-12 (c); and the intrinsic electric fields ($E_{int}$) of the A-$m$ (b) and the B-$m$ (d). The $E_{int}$ data in (b) and (d) can be well fitted with $g/m^{f}$, and extrapolated to 20 uc.}
  \end{figure}

Furthermore, we can estimate an effective electric field $E_{int}$ by calculating the slope of further-averaged electrostatic potential. In order to elucidate the evolution trend of the electronic structure with the thickness $m$ of the KTO layer, we present the intrinsic electric field $E_{int}$ of the KTO film as a function of the KTO thickness $m$ in Figs. 4(b) and 4(d) for the A-$m$ and B-$m$ heterostructures, respectively, where $E_{int}$ is determined by the slope of the further-averaged electrostatic potential as a smooth curve and its direction is from the (TaO$_2$)$^+$ to (KO)$^-$ end. The calculated results have two distinctive features. On one hand, the structural relaxation makes the absolute value $|E_{int}|$ become obviously lower because it modifies the polar discontinuity at the interfaces and the stress-induced polarization within the KTO layer. For the A-12 heterostructure in Fig. 4(a), $|E_{int}|$ reduces from $|E_{int}| = 5.3 \times 10^{-2}$  V/{\AA} without structural relaxation to $|E_{int}| = 4.7 \times 10^{-2}$  V/{\AA} with structural relaxation; and for the B-12 in Fig. 4(b), $|E_{int}|$ decreases from $|E_{int}| = 5.1 \times 10^{-2}$  V/{\AA} to $|E_{int}| = 4.4 \times 10^{-2}$  V/{\AA} accordingly.  On the other hand, for both the A-$m$ and B-$m$, the absolute value of intrinsic electric field, $|E_{int}|$, decreases with the thickness $m$ increasing. This trend can be explained by observing that the larger $m$ is, the smaller the polar discontinuity at the interfaces is made by structural relaxation. In detail, for the A-$m$ heterostructures in Fig. 4(b), $|E_{int}|$ with structural relaxation decreases with $m$, reducing from $|E_{int}| = 13.3 \times 10^{-2}$  V/{\AA} for $m=4$ to $|E_{int}| = 4.7 \times 10^{-2}$ V/{\AA} for $m=12$. For the B-$m$ in Fig. 4(d), $|E_{int}|$ with relaxation decreases from $|E_{int}| = 11.4 \times 10^{-2}$  V/{\AA} for $m=4$ to $|E_{int}| = 4.4 \times 10^{-2}$  V/{\AA} for $m=12$, which are all comparable to  electric fields detected from KTO films at the stoichiometric KTO/KMF$_3$ (M=Zn, Ni) superlattice\cite{2020oxyfluoride}.

It is very interesting that the intrinsic electric field ($E_{int}$) data for $m=4\sim 12$, as a function of the thickness of the KTO ($m$),  can be well fitted with a simple function with two parameters $g$ and $f$,
\begin{equation}
E_{int}=g/m^{f},
\end{equation}
for both of the A-$m$ and B-$m$ STO/KTO/BTO heterostructures. The fitting parameters are summarized in Table \ref{table1}. It is noticeable that for both of the types the absolute value of $g$ for the optimized case, $|g|$, is smaller than that of the case without relaxation, $|g_0|$, and accordingly the exponent $f$ is also smaller than $f_0$. We also extrapolate it to $m=20$ uc in Figs, 4(b) and 4(d). Thus, the expression (1) implies that $E_{int}$ will become little when $m$ is very large.
These mean that the intrinsic electric field originates from the interfacial effect, and then the electrostatic potential difference between the two interfaces will be saturated when the KTO thickness is large. These trends imply that the structural optimization (including interfacial relaxation) makes us avoid any divergence (polar catastrophe\cite{nakagawa2006some}) due to the polarity of the KTO layer.

\begin{table}[!h]
\caption{The fitting paramers $g$ and $f$ for the optimized A-$m$ and B-$m$ heterostructures, with corresponding $g_0$ and $f_0$ for the cases without relaxation.}\label{table1}
\begin{ruledtabular}
\begin{tabular}{ccccc}
Type &  $g$  & $f$ &  $g_0$ & $f_0$ \\ \hline
A     &  47.65  & 0.925  &  116.66  & 1.237 \\
B     &  -39.00  & 0.876  &  -123.35  & 1.275
\end{tabular}
\end{ruledtabular}
\end{table}

\begin{figure}[h]
\centering
\includegraphics[width=0.38\textwidth]{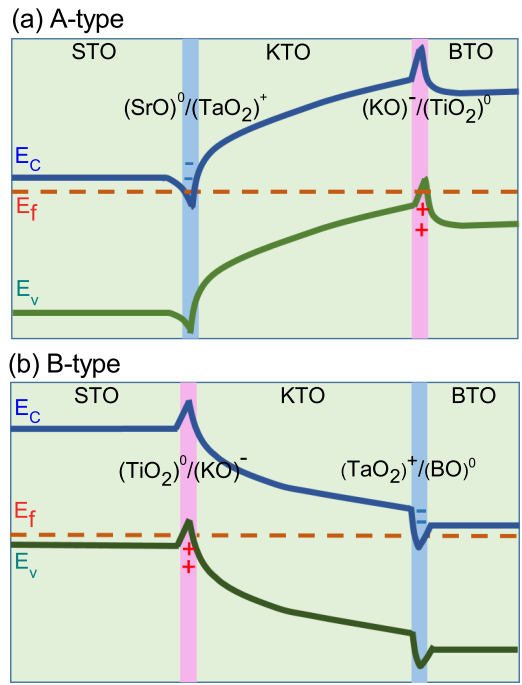}
\caption{\label{Fig5}
Schematic aligned band diagrams for the A-type (a) and B-type (b) STO/KTO/BTO heterostructures. The signs '+' and '-' indicate the 2DHG holes and 2DEG electrons at the interfaces.
}
\end{figure}

\subsection{Aligned band diagrams towards applications}

Combining all together, we show the aligned band diagrams of the A-type and B-type STO/KTO/BTO heterostructures in Fig. 5. The most standing-out feature is that there is a substantial energy band bending over the KTO layer and there exists a pair of highly-confined 2DEG and 2DHG at the interfaces for both of the heterostructures. The left-side STO/KTO [the right-side KTO/BTO] interface consists of (SrO)$^0$/(TaO$_2$)$^+$ [ (KO)$^-$/(TiO$_2$)$^0$ ] for the A-type heterostructure, and the interfaces consist of (TiO$_2$)$^0$/(KO)$^-$ and (TaO$_2$)$^+$/(BO)$^0$, respectively, for the B-type. It is remarkable that the band bending of the conduction band edge brings the conduction band on the KTO side below the Fermi level and that of the valence band edge brings the valence band on the BTO or STO side above the Fermi level. As we pointed out, the KTO thickness $m$ is a key parameter and with $m\ge m_c$ uc ($m_c=6$ uc as given above) there exist 2DEG and 2DHG at the interfaces for both A-type and B-type STO/KTO/BTO heterostructures. Our analysis indicates that the 2D carrier concentration ($n_{\rm 2DEG}=n_{\rm 2DHG}$) increases with $m$ monotonously, reaching a saturation value in the order of $10 ^{13}$ cm$^{-2}$ in the large-$m$ limit.

The band bending due to the intrinsic electric field can be attributed mainly to two origins, namely the polar discontinuity at the interfaces (the divergence is avoided effectively) \cite{nakagawa2006some} and the biaxial-stress-induced electric polarization within the KTO layer (the in-plane $a$ is set to $a_{\rm STO}$ and the z-axis lattice constant is optimized)\cite{Zhang2018Strain}. The polar discontinuity at the interface including (TaO$_2$)$^+$ causes a surface charge 0.5$e/a^2$ at most, and the polarization within the KTO layer can contribute at most -0.476$e/a^2$ at the interface, but actually they are both renormalized by the structural optimization and electron reconstruction. Consequently, for A-12 (B-12) the electron carrier concentration of 2DEG is 0.059$e/a^2$ (0.034$e/a^2$), in the order of the 'ideal' value 0.024$e/a^2$. It is similar at the corresponding hole-hosting interfaces and in the cases of other thicknesses. Naturally, the band bending can be manipulated by applying a gating field, and such an external field can also modify the critical thickness $m_c$ and the 2D carrier concentrations.

Because there are no dopants in both of the STO/KTO/BTO heterostructures, there is no defect scattering for the effective carriers of the 2DEG and 2DHG. Since the electron carriers of the 2DEG originate from Ta-5d$_{xy}$ and the hole carriers of the 2DHG from O-2p$_x$/p$_y$, inter-band scattering is also minimized. Furthermore, inter-carrier interaction should be small because the 2D carrier concentrations are in the order of $10 ^{13}$ cm$^{-2}$. Fortunately, these are favorable for enhancing the carrier mobility. It is expected that the 2D carrier concentrations can be controlled by applying a gating field, and thus these pairs of 2DEG and 2DHG could be useful for electronic and optoelectronic device applications such as novel transistors and sensors\cite{add4}. On the other hand, some spontaneous inter-interface excitons can be formed, with their life-time manipulated by applying a gating field or changing the thickness of the intermediate KTO layer, and interesting Bose-Eistein condensation could be observed in such exciton systems. If the 2DEG and 2DHG are both made superconducting\cite{2021two,chen2021electric}, some exciting effects and devices could be realized. Hopefully, these KTO-sandwiching systems could stimulate more exploration for new phenomena and novel devices.

\section{Conclusion}

In summary, through first-principles investigation and analysis, we have found two types of coexisting 2DEG and 2DHG at the interfaces in undoped STO/KTO/BTO heterostructures, with the surfaces insulating, when the KTO thickness $m$ reaches a crititcal value $m_c$. The two interfaces consist of (SrO)$^0$/(TaO$_2$)$^+$  and (KO)$^-$/(TiO$_2$)$^0$  for the A-type, and (TiO$_2$)$^0$/(KO)$^-$ and (TaO$_2$)$^+$/(BO)$^0$ for the B-type. The 2D electron carriers originate from Ta-$5 d_{xy}$ states at the interface including TaO$_2$ AL, and the hole carriers from O-$2 p_x/p_y$ orbitals at the other interface. The electron effective mass is 0.3$m_0$, and the hole effective mass ranges from 1.06$m_0$ to 1.12$m_0$. The carriers are highly confined at the corresponding interfaces, and the 2D carrier concentrations are in the order of $10 ^{13}$ cm$^{-2}$. The interfacial 2DEG and 2DHG are especially interesting because STO, KTO, and BTO bulks are insulating. Our analysis indicates that the polar discontinuity at the interfaces including the (TaO$_2$)$^+$ and (KO)$^-$ causes global structural optimization and partial electron reconstruction over the KTO layer in order to avoid electrostatic divergence, and on the other hand pins the direction of stress-induced electric polarization within the KTO layer, and consequently the 2DEG and 2DHG are simultaneously formed at the interfaces because of the band bending due to the intrinsic electric field effects in the KTO layer. These interfacial 2DEG and 2DHG could stimulate more exploration for more sandwich heterostructures, new phenomena, and novel devices.


\begin{acknowledgments}
This work is supported by the Nature Science Foundation of China (Grant Nos.11974393 and 11574366) and the Strategic Priority Research Program of the Chinese Academy of Sciences (Grant No. XDB33020100). All the numerical calculations were performed in the Milky Way \#2 Supercomputer system at the National Supercomputer Center of Guangzhou, Guangzhou, China.
\end{acknowledgments}


%

\end{document}